\documentclass[12pt]{article}

\usepackage{latexsym}

\usepackage{graphicx}

\begin{document}


\title{Generating  dyonic solutions in  5D Einstein-dilaton gravity with  antisymmetric forms  \\
and dyonic  black rings }

\author{
     Stoytcho S. Yazadjiev \thanks{E-mail: yazad@phys.uni-sofia.bg}\\
{\footnotesize  Department of Theoretical Physics,
                Faculty of Physics, Sofia University,}\\
{\footnotesize  5 James Bourchier Boulevard, Sofia~1164, Bulgaria }\\
}

\date{}

\maketitle

\begin{abstract}
We consider  $5$-dimensional Einstein-dilaton gravity with antisymmetric forms.
Assuming staticity and a restriction on the dilaton coupling parameters,
we derive $4$-dimensional sigma-model  with a target space  $SL(2,R)/SO(1,1)\times SL(2,R)/SO(1,1)$.
On this basis, using the symmetries of the target space, we develop a solution generating technique
and employ  it to construct new asymptotically flat and non-flat dyonic  black rings solutions.
The solutions are analyzed and the basic physical quantities are calculated.
\end{abstract}


\sloppy

\section{Introduction}

An interesting development in the black holes studies is the discovery of the black ring solutions
of the five-dimensional Einstein equations by Emparan and Reall \cite{ER1}, \cite{ER2}. These are
asymptotically flat solutions with an event horizon of topology $S^2\times S^1$ rather the
much more familiar $S^3$ topology. Moreover, it was shown in \cite{ER2} that both the black hole
and the the black ring can carry the same conserved charges, the mass and a single angular
momentum, and therefore there is no uniqueness theorem in five dimensions. Since the Emparan and
Reall's discovery many explicit examples of black ring solutions were found in various gravity
theories \cite{E}-\cite{P}. Elvang was able
to apply Hassan-Sen transformation to the solution \cite{ER2} to find a charged black ring in the
bosonic sector of the truncated heterotic string theory\cite{E}. This solution is the first
example of black rings with dipole charges depending, however, of the other physical parameters.
A supersymmetric black ring in
five-dimensional minimal supergravity was derived in \cite{EEMR1} and then generalized to the
case of concentric rings in \cite{GG1} and \cite{GG2}. A static black ring solution of the five
dimensional Einstein-Maxwell gravity was found by Ida and Uchida in \cite{IU}. In \cite{EMP}
Emparan derived "dipole black rings" in Einstein-Maxwell-dilaton (EMd) theory
in five dimensions. In this work Emparan showed that the black rings can exhibit novel
feature with respect to the black holes. The black rings can also carry independent nonconserved charges
which can be varied continuously without altering the conserved charges. This fact leads to
continuous non-uniqness. Following the same path that yields the
three-charge rotating black holes \cite{BLMPSV}-\cite{TSEY}, Elvang, Emparan and Figueras constructed a
seven-parameter family of supergravity solutions that describe non-supersymmetric black rings
and black tubes with three charges, three dipoles and two angular momenta \cite{EEF}.

The thermodynamics of the dipole black rings was studied first by Emparan in \cite{EMP} and by
Copsey  and Horowitz in \cite{CH}. Within the framework of the quasilocal
counterterm method, the thermodynamics of the dipole rings was discussed by Astefanesei and Radu \cite{AR}.
The first law of black rings thermodynamics in $n$-dimensional Einstein dilaton gravity with $(p+1)$-form field
strength was derived by Rogatko in \cite{ROG}. Static and asymptotically flat black rings solutions in five-dimensional
EMd gravity with arbitrary dilaton coupling parameter $\alpha$ were presented in \cite{KL}.
Non-asymptotically flat black rings immersed in external electromagnetic fields were found
and discussed  in \cite{O}, \cite{KL} and \cite{Y0}.A systematical derivation of the asymptotically
flat static black ring solutions
in five-dimensional EMd gravity with an arbitrary dilaton coupling parameter was given in
\cite{Y}. In the same paper and in \cite{Y1}, the author systematically derived new type static and rotating
black ring solutions which are not asymptotically flat.

In the present work we consider  $5$-dimensional Einstein-dilaton  gravity with antisymmetric forms. Assuming  staticity
we derive $4$-dimensional sigma-model  with a target space $SL(2,R)/SO(1,1)\times SL(2,R)/SO(1,1)$. On this basis,
we develop a solution generating technique and employ it to construct new dyonic black ring solutions. The solutions
are analyzed and the basic physical quantities are calculated.

\section{Basic equations and solution generating}

We consider the action

\begin{eqnarray}\label{A}
S= {1\over 16\pi} \int d^{5}x \sqrt{-g}\left[R - 2g^{\mu\nu}\partial_{\mu}\varphi\partial_{\nu}\varphi
- {1\over 4} e^{-2\alpha\varphi} F_{\mu\nu}F^{\mu\nu} - {1\over 12}e^{-2\beta\varphi} H^{\mu\nu\lambda}H_{\mu\nu\lambda}  \right]
\end{eqnarray}

where $H = dB$ and $B$ is the Kalb--Ramond field . This action is the 5-dimensional version of
the action studied by Gibbons and Maeda \cite{GM}. Let us note that the 3-form field strength $H$ can be dualized to 2-form field
strength ${\cal F}$ whose contribution to the action is given by $-{1\over 4}e^{2\beta\varphi}{\cal F}_{\mu\nu}{\cal F}^{\mu\nu}$. In other
words the theory we consider is equivalent to the Einstein-Maxwell-dilaton gravity with two distinct Maxwell fields and dilaton coupling
parameters. Particular examples of the action (\ref{A})(or its dual version)  with concrete values of the dilaton coupling
parameters arise from  string theory and supergravity via compactifications to five dimensions\footnote{As a result of the
compactifications, many additional fields come into play and one obtains rather complicated field models. In order
to obtain a simplified (truncated) action, as one we consider here, we must suppress the additional fields by imposing certain selfconsistent
conditions (see for example \cite{SV}- \cite{KK}).             }.

The action (\ref{A}) yields the following field equations

\begin{eqnarray}\label{FE1}
R_{\mu\nu} &=& 2\partial_{\mu}\varphi \partial_{\nu}\varphi
+ {1\over 2}e^{-2\alpha\varphi} \left(F_{\mu\lambda}F_{\nu}^{\,\lambda} - {1\over 6} F_{\sigma\lambda}F^{\sigma\lambda} g_{\mu\nu}\right)
 \nonumber \\ &+& {1\over 4}e^{-2\beta\varphi} \left(H_{\mu\sigma\lambda}H_{\nu}^{\,\sigma\lambda}
- {2\over 9}H_{\rho\sigma\lambda}H^{\rho\sigma\lambda}g_{\mu\nu} \right),\nonumber  \\
\nabla_{\mu}\nabla^{\mu}\varphi &=& - {\alpha \over 8} e^{-2\alpha\varphi} F_{\sigma\lambda}F^{\sigma\lambda} -
{\beta\over 24 }e^{-2\beta\varphi}H_{\rho\sigma\lambda}H^{\rho\sigma\lambda}, \\
&&\nabla_{\mu}\left(e^{-2\alpha\varphi} F^{\mu\nu} \right) = 0 , \nonumber \\
&&\nabla_{\mu}\left(e^{-2\beta\varphi} H^{\mu\nu\lambda} \right) = 0 . \nonumber
\end{eqnarray}

Here we will consider static solutions. The metric of static spacetimes can be written in the form

\begin{equation}
ds^2 = -e^{2u}dt^2 + e^{-u}h_{ij}dx^idx^j
\end{equation}

where $h_{ij}$  is four-dimensional metric with Euclidean signature.  We take the  fields $F$ and $H$ in the form

\begin{eqnarray}
F &=& - 2d\Phi \wedge dt ,\\
H &=& - 2e^{2\beta\varphi} \star (d\Psi \wedge dt).
\end{eqnarray}

Here $\star$ is the Hodge dual and the potentials $\Phi$ and $\Psi$ depend on the coordinates $x^{i}$ only.
With all these assumptions  the field equations are reduced to the following system

\begin{eqnarray}\label{DRFE}
&&{\cal D}_{i}{\cal D}^{i}u = {4\over 3} e^{-2u - 2\alpha\varphi}h^{ij} \partial_{i}\Phi \partial_{j}\Phi
+ {4\over 3} e^{-2u + 2\beta\varphi}h^{ij} \partial_{i}\Psi \partial_{j}\Psi ,\\
&&{\cal D}_{i}{\cal D}^{i}\varphi =  \alpha e^{-2u - 2\alpha\varphi}h^{ij} \partial_{i}\Phi \partial_{j}\Phi
- \beta e^{-2u + 2\beta\varphi}h^{ij} \partial_{i}\Psi \partial_{j}\Psi ,\\
&& {\cal D}_{i}\left(e^{-2u-2\alpha\varphi}{\cal D}^{i}\Phi \right) = 0,\\
&& {\cal D}_{i}\left(e^{-2u + 2\beta\varphi}{\cal D}^{i}\Psi \right) = 0,\\
&&R(h)_{ij} = {3\over 2}\partial_{i}u\partial_{j}u +  2\partial_{i}\varphi\partial_{j}\varphi
- 2e^{-2u -  2\alpha\varphi}\partial_{i}\Phi\partial_{j}\Phi - 2 e^{-2u +2\beta\varphi}\partial_{i} \Psi\partial_{j}\Psi ,
\end{eqnarray}

where ${\cal D}_{i}$ and $R(h)_{ij}$ are the  covariant derivative and the Ricci tensor with respect to the  metric $h_{ij}$.

These equations can be derived from the action

\begin{eqnarray}\label{DRA}
S = {1\over 16\pi} \int d^4x \sqrt{h} [R(h) - {3\over 2}h^{ij}\partial_{i}u\partial_{j}u - 2h^{ij}\partial_{i}\varphi \partial_{j}\varphi
 \nonumber \\ + 2e^{-2u -  2\alpha\varphi}h^{ij}\partial_{i}\Phi\partial_{j}\Phi + 2 e^{-2u +2\beta\varphi}h^{ij}\partial_{i}\Psi\partial_{j}\Psi  ].
\end{eqnarray}

It is convenient to introduce the rescaled potentials and parameters :

\begin{eqnarray}
\varphi_{*} &=& {2\over \sqrt{3}} \varphi , \,\,\,
\Phi_{*} = {2\over \sqrt{3}} \Phi , \,\,\,
\Psi_{*} = {2\over \sqrt{3}} \Psi , \\
\alpha_{*} &=& {\sqrt{3}\over 2} \alpha , \,\,\,
\beta_{*} = {\sqrt{3}\over 2} \beta .
\end{eqnarray}

It turns out that the action (\ref{DRA}) possesses an important group of  symmetries when
the coupling  parameters $\alpha$ and $\beta$ satisfy\footnote{It has to be noted that this condition fixes values of $\alpha$
(and $\beta$) diferent from those predicted by string theory.}

\begin{equation}
\alpha_{*} \beta_{*} = 1 .
\end{equation}

In order to see that, we define the new fields $\xi= u + \alpha_{*}\varphi_{*}$ and $\eta = u - \beta_{*}\varphi_{*}$
and introduce the symmetric matrices

\begin{eqnarray}
M_{1} = e^{-\xi}\left(%
\begin{array}{cc}
  e^{2\xi} - (1 + \alpha^2_{*})\Phi^2_{*} & - \sqrt{1 + \alpha^2_{*}}\Phi_{*} \\
- \sqrt{1 + \alpha^2_{*}}\Phi_{*}  &  -1 \\\end{array}%
\right) ,
\end{eqnarray}

\begin{eqnarray}
M_{2} = e^{-\eta}\left(%
\begin{array}{cc}
  e^{2\eta}- (1 + \beta^2_{*})\Psi^2_{*} & - \sqrt{1 + \beta^2_{*}}\Psi_{*} \\
- \sqrt{1 + \beta^2_{*}}\Psi_{*}  &  -1 \\\end{array}%
\right) ,
\end{eqnarray}

with $\det M_{1}= \det M_{2}=-1$ and $\beta_{*}=1/\alpha_{*}$.  Then the action (\ref{DRA})  can be written into the form

\begin{eqnarray}\label{SMA}
S = {1\over 16\pi} \int d^4x \sqrt{h} \left[R(h)
+ {3\over 4(1 + \alpha^2_{*})} h^{ij} Tr\left(\partial_{i}M_{1}\partial_{j}M^{-1}_{1}  \right)
\nonumber \right. \\
\left.+ {3\alpha^2_{*}\over 4(1 + \alpha^2_{*})} h^{ij} Tr\left(\partial_{i}M_{2}\partial_{j}M^{-1}_{2}  \right)\right] .
\end{eqnarray}

It is now clear that the action is invariant under the action of the group $SL(2,R)\times SL(2,R)$

\begin{eqnarray}
M_{1} \to G_{1}M_{1}G^{T}_{1} ,\,\,\, M_{2} \to G_{2}M_{2}G^{T}_{2} ,\,\,\,
\end{eqnarray}

where $G_{1}, G_{2} \in SL(2,R)$. The matrices $M_{1}$ and $M_{2}$ parameterize a coset
$SL(2,R)/SO(1,1)$. Therefore the action  corresponds to a non-linear $\sigma$-model action
with a target space $SL(2,R)/SO(1,1)\times SL(2,R)/SO(1,1)$. This is a generalization of $\sigma$-model studied
in  \cite{Y} and $\cite{Y1}$. As one should expect, the coset space is a product of two identical cosets which is a consequence
of the fact that our theory is equivalent to Einstein-Maxwell-dilaton gravity with two distinct Maxwell fields.

In the asymptotically flat case
\footnote{More precisely, we mean solutions with
"flat asymptotic" for the matrices $M_{1}$ and $M_{2}$, i.e. $M_{1}(\infty )=M_{2}(\infty )=\sigma_{3}$.},
without loss of generality we can set

\begin{equation}
u(\infty) = \varphi(\infty) = \Phi(\infty) = \Psi(\infty) = 0,
\end{equation}

i.e.

\begin{eqnarray}
M_{1}(\infty) = M_{2}(\infty) = \left(%
\begin{array}{cc}
  1 & 0 \\
 0 &-1 \\ \end{array}%
\right) = \sigma_{3}.
\end{eqnarray}

The transformations preserving the asymptotics are those satisfying

\begin{equation}
B_{1,2}\sigma_{3}B^{T}_{1,2} = \sigma_{3}.
\end{equation}

Therefore we find that $B_{1,2}\in SO(1,1)\subset SL(2,R)$.
Here we will adopt the  parameterization

\begin{eqnarray}
B_{1,2} = \left(%
\begin{array}{cc}
  \cosh\theta_{1,2} & \sinh\theta_{1,2} \\
 \sinh\theta_{1,2} &  \cosh\theta_{1,2} \\ \end{array}
 \right).
\end{eqnarray}

The symmetries of the dimensionally reduced field equations (\ref{DRFE}) can be employed
to generate new solutions from known ones, in particular, new solutions from known solutions of the vacuum Einstein equations.
Let us consider a static solution of the five-dimensional vacuum Einstein equations

\begin{eqnarray}\label{SES}
ds^2_{0} = - e^{2u_{0}}dt^2 + e^{-u_{0}}h_{ij}dx^idx^j
\end{eqnarray}

which is encoded into the matrices

\begin{eqnarray}
M^{(0)}_{1} = M^{(0)}_{2} = \left(%
\begin{array}{cc}
  e^{u_{0}} & 0 \\
 0 & -e^{-u_{0}} \\ \end{array}%
\right)
\end{eqnarray}

and the metric $h_{ij}$. The $SO(1,1)\times SO(1,1)$ transformations then generate new static
solutions to the filed equations (\ref{FE1}) given by the matrices

\begin{equation}
M_{1} = B_{1}M^{(0)}_{1}B^{T}_{1} , \,\,\, M_{2} = B_{2}M^{(0)}_{2}B^{T}_{2}
\end{equation}

and the same metric $h_{ij}$. In explicit form we have

\begin{eqnarray}\label{SGT}
e^{2u} &=& e^{2u_{0}}{\left(\cosh^2\theta_{1} - e^{2u_{0}}\sinh^2\theta_{1} \right)^{-2/(1+ \alpha^2_{*})}
\over
\left(\cosh^2\theta_{2} - e^{2u_{0}}\sinh^2\theta_{2} \right)^{2\alpha^2_{*}/(1+ \alpha^2_{*})} }, \nonumber \\
e^{-2\alpha_{*}\varphi_{*}} &=& \left[ { \cosh^2\theta_{1} - e^{2u_{0}}\sinh^2\theta_{1}
  \over \cosh^2\theta_{2} - e^{2u_{0}}\sinh^2\theta_{2} } \right]^{2\alpha^2_{*}/(1+ \alpha^2_{*}) } \\
\Phi_{*} &=& {\tanh\theta_{1} \over \sqrt{1+ \alpha^2_{*}} }   {1 - e^{2u_{0}} \over  1 - e^{2u_{0}}\tanh^2\theta_{1}} \nonumber , \\
\Psi_{*} &=&
{\alpha_{*}\tanh\theta_{2} \over \sqrt{1+ \alpha^2_{*}} }   {1 - e^{2u_{0}} \over  1 - e^{2u_{0}}\tanh^2\theta_{2}} \nonumber . \end{eqnarray}

In the next section we shall employ these solution generating formulas in order to obtain new black ring solutions
to the field equations (\ref{FE1}).

\section{Dyonic black rings}

As an explicit example of a seed vacuum Einstein solution we consider the five-dimensional,
static black ring solution given by the metric

\begin{eqnarray}\label{BRS}
ds^2_{0} &=& - {F(y)\over F(x)}dt^2 \\ &+& {{\cal R}^2\over (x-y)^2}
\left[F(x)(y^2-1)d\psi^2 + {F(x)\over F(y)} {dy^2\over (y^2-1)}
 + {dx^2\over (1-x^2) } + F(x) (1-x^2) d\phi^2\right] \nonumber
\end{eqnarray}

where
$F(x)= 1 + \lambda x$, ${\cal R}>0$ and $0<\lambda < 1$.
The coordinate $x$ is in the range $-1\le x\le 1$ and the
coordinate $y$ is in the range $-\infty<y\le -1$.
The solution has a horizon at $y=-1/\lambda$ . The topology of the horizon is $S^2\times S^1$,
parameterized by $(x,\phi)$ and $\psi$, respectively. In order to avoid a conical singularity
at $y=-1$ one must demand that the period of $\psi$
satisfies $\Delta \psi=2\pi/\sqrt{1 - \lambda}$. If one demands regularity at $x=-1$ the period of
$\phi$ must be $\Delta \phi=2\pi/ \sqrt{1 -\lambda}$. In this case the solution is
asymptotically flat and the ring is sitting on the rim of disk shaped membrane with
a negative deficit angle.
To enforce regularity at $x=1$ one must take $\Delta \phi = 2\pi/ \sqrt{1+\lambda}$
and the solution describes a black ring sitting on the rim of a disk shaped hole
in an infinitely extended deficit membrane with positive deficit.
More detailed analysis of the black ring solution can be found in \cite{ER1}.

Applying the transformation (\ref{SGT}) to the neutral black ring solution (\ref{BRS})
we obtain the following dyonic  solution

\begin{eqnarray}
ds^2  &=&   - {F(y)\over F(x)} { \left(\cosh^2\theta_{1} - {F(y)\over F(x) }\sinh^2\theta_{1} \right)^{- 2/(1+ \alpha^2_{*})} \over
\left(\cosh^2\theta_{2} - {F(y)\over F(x) } \sinh^2\theta_{2} \right)^{2\alpha^2_{*}/(1+ \alpha^2_{*})}} dt^2  \\
&+& { \left(\cosh^2\theta_{1} - {F(y)\over F(x) }\sinh^2\theta_{1} \right)^{1/(1+ \alpha^2_{*})}
 \over  \left(\cosh^2\theta_{2} - {F(y)\over F(x) } \sinh^2\theta_{2} \right)^{- \alpha^2_{*}/(1+ \alpha^2_{*})} }
\times  \nonumber \\
&& {{\cal R}^2\over (x-y)^2}
\left[F(x)(y^2-1)d\psi^2   + {F(x)\over F(y)} {dy^2\over (y^2-1)}
 + {dx^2\over (1-x^2) } + F(x) (1-x^2) d\phi^2\right] , \nonumber \\
e^{-2\alpha\varphi} &=& \left[ { \cosh^2\theta_{1} - {F(y)\over F(x) }\sinh^2\theta_{1}
  \over \cosh^2\theta_{2} - {F(y)\over F(x) }\sinh^2\theta_{2} } \right]^{2\alpha^2_{*}/(1+ \alpha^2_{*}) } ,\\
\Phi &=& {\sqrt{3}\tanh\theta_{1} \over 2\sqrt{1+ \alpha^2_{*}} }   {1 - {F(y)\over F(x) } \over  1 - {F(y)\over F(x) }\tanh^2\theta_{1}} , \\
\Psi &=& {\sqrt{3}\alpha_{*}\tanh\theta_{2} \over 2\sqrt{1+ \alpha^2_{*}} }   {1 - {F(y)\over F(x) } \over  1 - {F(y)\over F(x) }\tanh^2\theta_{2}} .
\end{eqnarray}

It is worth noting that this new solution is a generalization of the solution of \cite{KL} which can be obtained as a particular case
by setting $\theta_{2}=0$.   The analysis of our dyonic solution is  quite similar to that for the neutral black ring \cite{ER1}.
From the explicit form of the metric it is clear that there is a horizon at $y= - 1/\lambda$. Using arguments
similar to those in \cite{ER1} one can show that the horizon is regular. The metric of a constant $t$ slice
through the horizon is

\begin{eqnarray}\label{HM}
ds^2_{h} = [\cosh^2\theta_{1}]^{1/(1 + \alpha^2_{*})} [\cosh^2\theta_{2}]^{\alpha^2_{*}/(1 + \alpha^2_{*})} {{\cal R}^2 \lambda^2
\over F^2(x)} \times \nonumber \\ \left[F(x){1-\lambda^2\over \lambda^2 } d\psi^2 + {dx^2 \over 1-x^2} + F(x)(1-x^2)d\phi^2 \right] .
\end{eqnarray}

To analyze the case when  $y\to -1$, we set $y=-\cosh\rho$. Near $\rho=0$ the $ty$ part of the metric is
conformal to

\begin{equation}
dl^2_{ty} \approx d\rho^2 + (1-\lambda)\rho^2 d\psi^2.
\end{equation}

This is regular at $\rho=0$ provided that $\psi$ is identified with period $\Delta \psi = 2\pi/\sqrt{1-\lambda}$.
Let us consider now the $x\phi$ part of the metric which is conformal to

\begin{equation}
dl^2_{x\phi} \approx {dx^2 \over 1-x^2 } + F(x)(1-x^2)d\phi^2.
\end{equation}

In order for this metric to be regular at $x=-1$ we must identify $\phi$ with period $\Delta \phi= 2\pi/\sqrt{1- \lambda}$.
The regularity at $x=1$ requires to impose the period $\Delta=2\pi/\sqrt{1+\lambda}$. It is therefore
not possible to have regularity at both $x=-1$ and $x=1$. The regularity at $x=-1$ means the presence of
a conical singularity at $x=1$ and vice versa. In both cases the $x\phi$ part of the metric describes a surface that
is topologically $S^2$ with a conical singularities at one of the poles. The above analysis show that the horizon
metric (\ref{HM}) describes a hypersurface with topology $S^2\times S^1$. In addition, there is an inner spacelike
singularity   at  $y=-\infty$.

Using the explicit form of the horizon metric (\ref{HM}) one can calculate the area of the horizon

\begin{equation}
{\cal A}^{\pm}_{h} = {8\pi^2\lambda^2{\cal R}^3\over (1-\lambda)\sqrt{(1+\lambda)(1\pm \lambda)} }
[\cosh^3\theta_{1}]^{1/(1+ \alpha^2_{*})} [\cosh^3\theta_{2}]^{\alpha^2_{*}/(1+ \alpha^2_{*})}.
\end{equation}

Here the sign $+$ corresponds to the case of a conical singularity at $x=-1$ and vice versa.

The temperature can be found by Euclideanizing the metric and the result is

\begin{equation}
T = {\sqrt{1-\lambda^2} \over 4\pi {\cal R} \lambda }
[\cosh^3\theta_{1}]^{-1/(1+ \alpha^2_{*})} [\cosh^3\theta_{2}]^{-\alpha^2_{*}/(1+ \alpha^2_{*})} .
\end{equation}

In order to compute the mass of the dyonic solution we use the quasilocal formalism. After a  long algebra we
find

\begin{equation}
M^{\pm} = {3\pi\lambda {\cal R}^2\over 4 \sqrt{(1-\lambda)(1\pm \lambda) } }
[1 + {2\over 1 + \alpha^2_{*} }\sinh^2\theta_{1} + {2\alpha^2_{*}\over 1 + \alpha^2_{*} }\sinh^2\theta_{2}].
 \end{equation}

The electric and the magnetic charge are defined by

\begin{eqnarray}
Q &=& {1\over 8\pi } \oint_{S_{\infty}^{3}} \star  \, e^{-2\alpha\varphi} F ,\\
P &=& {1\over 8\pi } \oint_{S_{\infty}^{3}} H  .
\end{eqnarray}

After some algebra we find

\begin{eqnarray}
Q^{\pm} &=& {\sqrt{3} \sinh\theta_{1} \cosh\theta_{1}\over \sqrt{1+ \alpha^2_{*}} }
{\pi \lambda {\cal R}^2 \over \sqrt{(1-\lambda )(1\pm \lambda)} } ,\\
P^{\pm} &=& {\sqrt{3}\alpha_{*} \sinh\theta_{2} \cosh\theta_{2}\over \sqrt{1+ \alpha^2_{*}} }
{\pi \lambda {\cal R}^2 \over \sqrt{(1-\lambda )(1\pm \lambda)} } .
\end{eqnarray}

It is straightforward to check that the following Smarr-like relation is satisfied

\begin{equation}
M^{\pm} = {3\over 8} T{\cal A}_{h}^{\pm} + \Phi_{h}Q^{\pm} + \Psi_{h}P^{\pm}
\end{equation}

where subscript $h$ shows that the corresponding quantity is evaluated on the horizon.

\section{Dyonic black hole limit}

Here we consider the asymptotically flat dyonic black ring solution for which the conical singularity is
at $x=1$. Let us introduce the new parameter

\begin{equation}
m = {2 {\cal R}^2\over 1-\lambda }
\end{equation}

such that it remains finite as $\lambda \to 1$ and ${\cal R} \to 0$. Also, let us change the coordinates

\begin{eqnarray}\label{CH}
x &=& -1 + {2{\cal R}^2\sin^2\theta\over r^2 - m\sin^2\theta },  \\
y &=& -1 - {2{\cal R}^2\cos^2\theta\over r^2 - m\sin^2\theta },\nonumber
\end{eqnarray}

and rescale $\psi$ and $\phi$

\begin{equation}\label{CR}
(\psi,\phi) \to \sqrt{m\over 2{\cal R}^2} (\psi,\phi)
\end{equation}

so that they now have canonical periodicity $2\pi$. Then we obtain the solution

\begin{eqnarray} \label{AFDBHS}
ds^2 &=& - \left(1 -{m\over r^2}\right) {\left(1 + {m\over r^2}\sinh^2\theta_{1} \right)^{- 2/(1+ \alpha^2_{*})} \over
\left(1 + {m\over r^2}\sinh^2\theta_{2} \right)^{ 2\alpha^2_{*}/(1+ \alpha^2_{*})}} dt^2 \nonumber \\
&+& {\left(1 + {m\over r^2}\sinh^2\theta_{1} \right)^{1/(1+ \alpha^2_{*})} \over
\left(1 + {m\over r^2}\sinh^2\theta_{2} \right)^{ -\alpha^2_{*}/(1+ \alpha^2_{*})}}
\left[{dr^2\over 1 - {m\over r^2} } + r^2d\theta^2 + r^2\cos^2\theta d\psi^2 + r^2\sin^2\theta d\phi^2 \right],
\nonumber \\
e^{-2\alpha\varphi} &=&
\left[ {1 + {m\over r^2}\sinh^2\theta_{1}\over 1 + {m\over r^2}\sinh^2\theta_{2}}\right]^{2\alpha^2_{*}/ (1 + \alpha^2_{*})} ,\\
\Phi &=& {\sqrt{3}\sinh\theta_{1}\cosh\theta_{1}\over 2 \sqrt{1 + \alpha^2_{*}} } {m\over r^2 + m\sinh^2\theta_{1} }, \nonumber  \\
\Psi &=& {\sqrt{3}\alpha_{*}\sinh\theta_{2}\cosh\theta_{2}\over 2 \sqrt{1 + \alpha^2_{*}} } {m\over r^2 + m\sinh^2\theta_{2}}.
\nonumber
\end{eqnarray}

This solution describes a $5$-dimensional asymptotically flat dyonic black hole with a horizon at $r=m$ with usual $S^{\,3}$-topology.
There is also a regular inner horizon at $r=0$ with the same topology.
It is worth noting that the solution (\ref{AFDBHS}) can be obtained via the above discussed solution generating technique from the
$5$-dimensional Schwarzschild-Tangherlini black hole solution

\begin{equation}\label{STBH}
ds^2_{0} = - \left(1 -{m\over r^2}\right)dt^2 + {dr^2\over 1 - {m\over r^2} } + r^2d\theta^2 + r^2\cos^2\theta d\psi^2 + r^2\sin^2\theta d\phi^2
\end{equation}

with mass $M= {3\pi \over 8}m$. Our dyonic black hole solution is in fact the 5-dimensional
Gibbons-Maeda dyonic black hole\footnote{Note that the $5D$ Gibbons-Maeda solution is written here in coordinates different from those
used in \cite{GM}. } derived  via a completely different method in  \cite{GM}.

The physical quantities characterizing the dyonic black hole solution can be found as a limit of the
corresponding quantities for the black ring. Also, it can be checked that a Smarr-like relation is satisfied.

\section{Asymptotically non-flat dyonic black rings}

In this section we consider more exotic black ring solutions - black rings with unusual asymptotic.
These are dyonic black rings which are asymptotically non-flat. Such black hole and black brane solutions, although unusual, have attracted much interest  \cite{PW}-\cite{CGLO}.

In order to generate asymptotically non-flat dyonic black  ring solutions we consider
the special $SL(2,R)\times SL(2,R)$ transformation\footnote{The dual transformation $B\times N$ generate a dual
solution which can be also obtained via the "discrete duality"
$\varphi\to -\varphi$, $\Phi\to\Psi$ and $\alpha\to\beta$. In addition, we can consider the
transformation $N(a)\times N(b)$, too. This transformation, however, generates configurations with frozen (constant) dilaton field.
Such configurations will not be considered here. } given by  $N\times B$ where

\begin{eqnarray}
N = \left(%
\begin{array}{cc}
  0 & - a^{-1} \\
  a &  a \\ \end{array}
 \right) ,
\end{eqnarray}

\begin{eqnarray}
B = \left(%
\begin{array}{cc}
  \cosh\theta & \sinh\theta \\
 \sinh\theta &  \cosh\theta \\ \end{array}
 \right).
\end{eqnarray}

This transformation, applied to the seed solution (\ref{SES}),  generates new dyonic solutions with
the following potentials

\begin{eqnarray}
e^{2u} &=& e^{2u_{0}} { \left(\cosh^2\theta - e^{2u_{0}}\sinh^2\theta \right)^{-2\alpha^2_{*}/(1+ \alpha^2_{*})} \over
 \left[a^2 (1 - e^{2u_{0}}) \right]^{2/(1+ \alpha^2_{*})} } ,\\
e^{-2\alpha_{*}\varphi_{*}} &=&
\left[ a^2(1-e^{2u_{0}}) \over \cosh^2\theta - e^{2u_{0}}\sinh^2\theta  \right]^{2\alpha^2_{*}/(1+ \alpha^2_{*})} ,\\
\Phi_{*} &=& {1\over \sqrt{1+ \alpha^2_{*}} } {a^{-2}\over 1-e^{2u_{0}} } ,\\
\Psi_{*} &=&  {\alpha_{*}\tanh\theta\over \sqrt{1 + \alpha^2_{*}}} {1- e^{2u_{0}}\over  1 - e^{2u_{0}}\tanh^2\theta }.
\end{eqnarray}

Applying these formulas to the neutral black ring solution we obtain the following dyonic solution

\begin{eqnarray}
ds^2 &=& -{F(y)\over F(x)} { \left[ \cosh^2\theta  - {F(y)\over F(x)} \sinh^2\theta \right]^{-2\alpha^2_{*}/(1 +\alpha_{*}^2) }  \over
 \left[a^2(1 - {F(y)\over F(x) } ) \right]^{2/(1+ \alpha^2_{*})}  }  dt^2
\nonumber \\ &+& \left[ \cosh^2\theta  - {F(y)\over F(x)} \sinh^2\theta \right]^{\alpha^2_{*}/(1 +\alpha_{*}^2) }
\left[a^2(1 - {F(y)\over F(x) } ) \right]^{1/(1+ \alpha^2_{*})} \times  \\
&& {{\cal R}^2\over (x-y)^2}
\left[F(x)(y^2-1)d\psi^2   + {F(x)\over F(y)} {dy^2\over (y^2-1)}
 + {dx^2\over (1-x^2) } + F(x) (1-x^2) d\phi^2\right], \nonumber \\
e^{-2\alpha \varphi} &=& \left[ a^2(1 - {F(y)\over F(x) } ) \over  \cosh^2\theta  - {F(y)\over F(x)} \sinh^2\theta \right]^{2\alpha^2_{*}/(1+ \alpha^2_{*})} ,\\
\Phi &=& {\sqrt{3}\over 2 } {a^{-2}\over \sqrt{1+ \alpha^{2}_{*}}} {1\over 1 - {F(y)\over F(x) }} ,\\
\Psi &=& {\sqrt{3}\over 2 } {\alpha_{*}\tanh\theta\over \sqrt{1+ \alpha^{2}_{*}}} {1 - {F(y)\over F(x)} \over 1 - {F(y)\over F(x)}\tanh^2\theta}
\end{eqnarray}

The metric has a horizon at $y=-1/\lambda $. Arguments similar to those of \cite{ER1} show
that the horizon is regular. Further analysis of the solution is quite similar to the asymptotically
flat case. That is why we present only the results. In order to get rid of the possible conical singularity
at $y=-1$ we must identify $\psi$ with a period $\Delta= 2\pi/\sqrt{1-\lambda}$. It is not possible to have
regularity at $x=1$ and $x=-1$. The regularity at $x=1$ requires the period $\Delta \phi= 2\pi/\sqrt{1+\lambda}$
while to avoid the conical singularity at $x=-1$ we must impose $\Delta \phi= 2\pi/\sqrt{1-\lambda}$.  Also,
the solution has an inner spacelike singularity at $y=-\infty$.

In order to see the asymptotic behavior of the solution at spacial infinity ($x=y=-1$) let us introduce the
new coordinates

\begin{equation}
r\cos\theta = {\cal R} {\sqrt{y^2 -1}\over x-y },\,\,\,
r\sin\theta = {\cal R} {\sqrt{1-x^2}\over x-y },\,\,\,
{\tilde \psi} = \sqrt{1-\lambda}\psi ,\,\,\, {\tilde \phi}= \sqrt{1- \lambda}\phi.
\end{equation}

Then for $r\to \infty$ we find

\begin{eqnarray} \label{SBRAM}
ds^2 &\approx& - \left[{(1-\lambda)\over 2\lambda } {r^2\over a^2{\cal R}^2}\right]^{2\over 1+\alpha^2_{*} } dt^2
\\
&+& \left[{(1-\lambda)\over 2\lambda }
 {r^2\over a^2{\cal R}^2}\right]^{-1\over 1+\alpha^2_{*} } \left[dr^2 +
 r^2d\theta^2 + r^2\cos^2\theta d{\tilde \psi}^2  + r^2\sin^2\theta d{\tilde \phi}^2  \right] ,\nonumber \\
e^{2\alpha\varphi} &\approx &\left[{(1-\lambda)\over 2\lambda }
{r^2\over a^2{\cal R}^2}\right] ^{2\alpha^2_{*}\over 1+\alpha^2_{*} } ,  \\
\Phi &\approx& {\sqrt{3}\over 2 \sqrt{1+ \alpha^2_{*}} }
\left[{(1-\lambda)\over 2\lambda } {r^2\over a^2{\cal R}^2}\right] ,\\
\Psi &\approx& {\sqrt{3}\alpha_{*}\sinh\theta\cosh\theta\over 2\sqrt{1+ \alpha^2_{*}} } {2\lambda\over 1-\lambda}
{{\cal R}^2\over r^2 }.
\end{eqnarray}

Even though our solution is not asymptotically flat and the dilaton field behaves like $\varphi \sim \ln(r)$
for large $r$,  the Ricci and the Kretschmann scalars vanish as $r\to \infty$. More precisely we have

\begin{equation}
R \sim  r^{-{2\alpha^2_{*} \over 1 + \alpha^2_{*}}}  ,\,\,\,\,
R_{\mu\nu\alpha\beta}R^{\mu\nu\alpha\beta}\sim r^{-{4\alpha^2_{*} \over 1 + \alpha^2_{*}} }.
\end{equation}

Therefore the spacetime is well-behaved for $r\to \infty$.

The temperature can be found via the surface gravity and the result is

\begin{equation}
T = {\sqrt{1-\lambda^2}\over  4\pi {\cal R}\lambda} [\cosh\theta]^{- 3\alpha^2_{*}/(1+ \alpha^2_{*}) }
a^{- 3/(1+ \alpha^2_{*})} .
\end{equation}

The horizon area is found by straightforward integration

\begin{equation}
{\cal A}^{\pm}_{h} = {8\pi^2 {\cal R}^3 \lambda^2\over (1-\lambda)\sqrt{(1+\lambda)(1\pm \lambda)}  }
[\cosh\theta]^{ 3\alpha^2_{*}/(1+ \alpha^2_{*}) }
a^{3/(1+ \alpha^2_{*})} .
\end{equation}

For the electric and the magnetic charge we obtain

\begin{eqnarray}
Q^{\pm} &=& {\sqrt{3} a^2\over \sqrt{1 + \alpha^2_{*}} } {\pi {\cal R}^2 \lambda \over \sqrt{(1-\lambda)(1\pm \lambda)} } ,\\
P^{\pm} &=& {\sqrt{3}\alpha_{*}\sinh\theta\cosh\theta \over \sqrt{1 + \alpha^2_{*}} } {\pi {\cal R}^2 \lambda \over \sqrt{(1-\lambda)(1\pm \lambda)} }.
\end{eqnarray}

The mass of the solution is calculated through the use of the quasilocal formalism. After a long algebra we find

\begin{equation}
M^{\pm} = {3\pi {\cal R}^2 \lambda \over 4 \sqrt{(1-\lambda)(1\pm \lambda)}}
\left[{\alpha^2_{*} \over 1 + \alpha^2_{*} } + {2\alpha^2_{*}\over 1 + \alpha^2_{*} }\sinh^2\theta \right].
\end{equation}

The electric potential $\Phi$ is defined up to an arbitrary additive constant.
In the asymptotically flat case there is a preferred gauge in which $\Phi(\infty)=0$. In the non-asymptotically
flat case the electric potential diverges at spacial infinity and there is no preferred gauge.
The arbitrary constant, however, can be fixed so that
the Smarr-type relation is  satisfied:

\begin{equation}
M^{\pm}= {3\over 8}T A^{\pm}_{h} + \Phi_{h}Q^{\pm} + \Psi_{h}P^{\pm}.
\end{equation}

\section{Asymptotically non-flat dyonic black hole limit}

Here we consider the dyonic black ring solution with a conical singularity at $x=1$.
Performing the coordinate change (\ref{CH}), the coordinate rescaling (\ref{CR}) and
taking the limit ${\cal R}\to 0$ and $\lambda \to 1$  with $m= {2{\cal R}^2/(1-\lambda)}$ fixed,
we obtain the following asymptotically non-flat dyonic black hole solution

\begin{eqnarray}\label{ANFDBHS}
ds^2 &=& -\left(1- {m\over r^2}\right)\left({r^2\over ma^2 }\right)^{2/(1+ \alpha^2_{*})}
\left(1 + {m\over r^2}\sinh^2\theta \right)^{-2\alpha^2_{*}/(1+ \alpha^2_{*})}dt^2 \nonumber \\
&+& \left(1 + {m\over r^2}\sinh^2\theta \right)^{\alpha^2_{*}/(1+ \alpha^2_{*})}
\left({r^2\over ma^2 }\right)^{-1/(1+ \alpha^2_{*})} \times \nonumber
\\ && \left[{dr^2\over 1 - {m\over r^2} } + r^2d\theta^2 + r^2\cos^2\theta d\psi^2 + r^2\sin^2\theta d\phi^2  \right], \nonumber  \\
e^{-2\alpha\varphi} &=& \left[{ma^2\over r^2 + m\sinh^2\theta }\right]^{2\alpha^2_{*}/(1+ \alpha^2_{*})}, \\
\Phi &=& {\sqrt{3}\over 2\sqrt{1+ \alpha^2_{*}} }  {r^2\over ma^2 } ,\nonumber \\
\Psi &=& {\sqrt{3}\alpha_{*}\sinh\theta\cosh\theta\over 2\sqrt{1+ \alpha^2_{*}} } {m\over r^2 + m\sinh^2\theta  }.\nonumber
\end{eqnarray}

This solution describes a $5$-dimensional asymptotically non-flat dyonic black hole  with a horizon at $r=m$ with usual $S^{\,3}$-topology.
There is also a regular inner horizon at $r=0$ with the same topology. It is worth noting that the solution (\ref{ANFDBHS}) can be obtained
via the above discussed solution generating technique from the $5$-dimensional Schwarzschild-Tangherlini black hole solution (\ref{STBH}).
The physical quantities characterizing the dyonic black hole solution can be found as a limit of the
corresponding quantities for the black ring. In addition, it can be checked, that the mass, the charges and
the horizon potentials satisfy a Smarr-like relation.

\section{Conclusion}

In this work we considered static $5$-dimensional Einstein-dilaton  gravity with antisymmetric forms.
We derived a $4$-dimensional $\sigma$-model with a target space
$SL(2,R)/SO(1,1)\times SL(2,R)/SO(1,1)$. Using the target space symmetries we constructed new
asymptotically  flat and non-flat dyonic black rings solutions. These solutions were analyzed and the basic physical
quantities were calculated.  It was shown that Smarr-type relations are satisfied.

The solution generating technique presented in this paper can be used for generating not only black ring and black hole
solutions but for generating  many others exact solutions. It is worth noting that the formulation of the
static equations in terms of a non-linear $\sigma$-model enables us to apply the integrable system techniques when
the target space variables depend on two coordinates only (i.e. in the presence of two additional commuting Killing
vectors). In this connection we should mention the related works on soliton solutions in 5-dimensional general
relativity \cite{MI},\cite{K},\cite{AK} and in some 5-dimensional low-energy string theories \cite{YU} - \cite{HAPTV}.

Finally, it would be interesting to find rotating dyonic black
rings. The construction of dyonic solutions with rotation is now in progress and the results
will be presented elsewhere.

\section*{Acknowledgements}

This work was partially supported by the Bulgarian National Science Fund under Grant MUF04/05 (MU 408)
and the Sofia University Research Fund under Grant N60.

\end{document}